\documentstyle[aps,epsfig]{revtex}
\begin{document}
\twocolumn[\hsize\textwidth\columnwidth\hsize
           \csname @twocolumnfalse\endcsname]
\title{Medium modifications of kaons in pion matter}
\author{B.V. Martemyanov$^{a,b)}$, Amand Faessler$^{b)}$, C. Fuchs$^{b)}$ and M.I.
Krivoruchenko$^{a,b)}$}
\address{$^{a)}${\it Institute for Theoretical and Experimental 
Physics, B. Cheremushkinskaya 25}\\
{\it 117259 Moscow, Russia}\\
$^{b)}${\it Institut f\"{u}r Theoretische Physik, 
Universit\"{a}t T\"{u}bingen, Auf der Morgenstelle 14}\\
{\it D-72076 T\"{u}bingen, Germany}}
\maketitle

\begin{abstract}
Kaon in-medium masses and mean-field potentials are calculated 
in isotopically symmetric pion matter to one loop of chiral 
perturbation theory. The results are extended to RHIC temperatures 
using experimental data on  $\pi K$ scattering phase shifts. 
The kaon in-medium broadening results in an acceleration of 
the $\phi \rightarrow K\bar{K}$ decay. The increased apparent 
dilepton branching of the $\phi $-mesons, observed recently by NA50, 
NA49, and the PHENIX collaborations at RHIC, is interpreted in terms 
of rescattering of secondary kaons inside of the pion matter.\\
{\bf keywords}: kaons, pions, heavy-ion collisions
\end{abstract}

The theoretical study of kaon properties in a dense nuclear
environment has a long history \cite{Kaplan,Brown,Waas}. To search for 
signals of predicted in-medium modifications at high baryon densities 
is one of the primary goals of the experiments devoted to the 
measurements of kaonic observables in heavy-ion reactions at
intermediate energies \cite{kaos}. Corresponding transport
calculations revealed significant evidence for in-medium modifications 
of the kaon properties during the course of such
reactions \cite{texas,tuebingen}. This picture was recently
complemented by the measurements of $K^{+}$ production in proton-nucleus 
reactions \cite{anke}.

In heavy-ion reactions at RHIC energies, the physical conditions are quite
different. The medium is baryon dilute but meson rich \cite{bravina02}.
Since by far the most abundant particles are pions, one can speak about pion
matter created in heavy-ion collisions at RHIC energies. It is therefore
interesting to consider the modifications of kaon properties in such a
pion-dominated medium. This question is also important from another point
of view: The $\phi $-meson is considered as a promising mesonic probe for
the fireball formed in such reactions \cite{KMR86,Sh85}.

Possible modifications of the $\phi $-meson dilepton branching in a pion gas
have been discussed one decade ago by Lissauer and Shuryak \cite{shur} and
by Blaizot and Galain \cite{biaz}. The $\mu ^{+}\mu ^{-}$ yield from $\phi $%
-meson decays was recently measured in central {\rm Pb+Pb} collisions at
CERN/SPS energies by the NA50 Collaboration \cite{NA50} and the $K\bar{K}$
yield was measured by the NA49 Collaboration \cite{NA49}. The number
of $\phi $-mesons detected through the dilepton channel was found to be a factor
of two to four greater than that of $\phi $-mesons detected trough the
kaon channel. This difference might be attributed to in-medium $\phi $ and $K$
mass shifts \cite{biaz} and/or rescattering of the secondary kaons in 
hadronic matter \cite{JJD01,Soff01}. Recently, preliminary $\phi $%
-meson$\ $production data in ${\rm Au+Au}$ collisions at RHIC energies were
reported by the PHENIX Collaboration \cite{phenix}. In this experiment,
the $\phi $ yield was simultaneously measured in the $\phi \rightarrow
e^{+}e^{-} $ and $K\bar{K}$ channels. The result is consistent with an
increased dilepton branching.

The kaon-pion gas was studied in details within chiral perturbation theory
(ChPT) in Refs. \cite{dobado,pelaez}, focusing thereby on the quark
condensates. In this letter, we discuss modifications of the kaon self-energy operator 
$\Sigma(p^2,E)$ and of the $\phi
\rightarrow K\bar{K}$ decay mode inside hot pion matter. ChPT, proposed
for description of interactions of pseudoscalar mesons at low energies, is an
adequate tool for studying the problem at low temperatures and 
useful to control the low temperature limit of phenomenological models.

The in-medium mass operator of kaons, $\Sigma (p ^{2},E)$, can be
expressed in terms of the $\pi K$ forward scattering amplitudes for on-shell
pions and off-shell kaons. The on-shell $\pi K$ amplitudes have been
calculated in ChPT to the order $p^{4}$ by several authors (see e.g. \cite{pik}
and references therein). To lowest order $p^{2}$, the off-shell $\pi K$
amplitudes are given in Ref. \cite{Weinberg}. 
Near the threshold, the isospin-even and odd $\pi K$ scattering amplitudes
can be written as 
\begin{eqnarray}
A^{\pm }(s,t,p^2)&=& 8\pi \sqrt{s}\left( a_{0}^{\pm }+p^{*2}(b_{0}^{\pm
}+3a_{1}^{\pm })
+\frac{3}{2}ta_{1}^{\pm }\right) \nonumber \\
&+& c^{\pm
}(p ^{2}-M_{K}^{2})  \label{amplitude}
\end{eqnarray}
where $a_{\ell }^{\pm }$ and $b_{\ell }^{\pm }$ are the $\pi K$ scattering
lengths and effective ranges, $p^{*}=p^{*}(\sqrt{s},M_{\pi },M_{K})$ is the
c.m. momentum of the $\pi K$ system, $s=(p + p_{\pi } )^{2},$ $t$ $%
=(p_{\pi }^{\prime }-p_{\pi })^{2},$ $p=(E,{\bf p})$ is the kaon momentum 
and $p ^{2}\neq
M_{K}^{2}$ in general.

The number densities of pions are given by Bose distributions 
\[
dn_{v\pi }=\frac{d^{3}p_{\pi }}{(2\pi )^{3}}\left( {\rm exp}(\frac{E_{\pi
}-\mu _{\pi }}{T})-1\right) ^{-1} 
\]
where $\mu _{\pi ^{+}}=-\mu _{\pi ^{-}}\ $is the $\pi ^{+}$ chemical
potential, $\mu _{\pi ^{0}}=0$. The scalar pion density is defined by $%
dn_{s\pi }=dn_{v\pi }/(2E_{\pi }).$ We assume $-M_{\pi }<\mu _{\pi
^{+}}<M_{\pi },$ so that a pion Bose condensate is not formed. 
Thermal properties of interacting pseudoscalar mesons ($\pi $, $K$%
, $...$) and nucleons are studied in Ref. \cite{plan}.

The sum of the forward $\pi ^{+}K$, $\pi ^{0}K$, and $\pi ^{-}K$ scattering
amplitudes can be integrated over the pion momenta in the rest frame of the
medium to obtain the $K$-meson self-energy operator: 
\begin{eqnarray}
-\Sigma (p ^{2},E)&=&\int A^{+}(s,0,p^2)(dn_{s\pi ^{+}}+dn_{s\pi ^{0}}+dn_{s\pi
^{-}})
\nonumber \\
&+&\int A^{-}(s,0,p^2)(-dn_{s\pi ^{+}}+dn_{s\pi ^{-}})~~.  \label{SE}
\end{eqnarray}
Near threshold the amplitudes $A^{\pm }(s,t,p^2)$ can be expanded up
to $\sim O(s-(M_{\pi }+M_{K})^{2}) + O(p^2 - M_{K})^{2})$, 
in which case the self-energy operator can be
written as follows 
\begin{equation}
-\Sigma (p ^{2},E)=(p ^{2}-M_{K}^{2})(Z_{K}^{-1}-1)-\delta
M_{K}^{2}-2EV_{K}.  
\label{expand}
\end{equation}
This representation allows to identify $\delta M_{K}$ as a mass shift and $%
V_{K}$ as an external vector potential. The poles of the propagator
appearing at $p ^{2}-M_{K}^{2}-\Sigma (p ^{2},E)=0$ determine the
in-medium kaon dispersion law 
\begin{equation}
E_{K}^{(\pm )}({\bf p})=\pm \sqrt{{\bf p}{}^{2}+M_{K}^{2}+\delta M_{K}^{2}{}}%
+V_{K}~~.  
\label{dl}
\end{equation}

Using current algebra predictions for the threshold $\pi K$-scattering
parameters \cite{Weinberg}, one gets 
\begin{eqnarray}
Z_{K^{\pm }}^{[0]-1} &=&1+\frac{n_{s\pi ^{+}}+n_{s\pi ^{0}}+n_{s\pi ^{-}}}{%
2F^{2}}~~,  
\label{Z0} \\
\delta M_{K^{\pm }}^{[0]2} &=&0~~,  
\label{dM0} \\
V_{K^{\pm }}^{[0]} &=&\pm \frac{n_{v\pi ^{+}}-n_{v\pi ^{-}}}{4F^{2}}~~
\label{V0}
\end{eqnarray}
where $F = 92$ MeV is the pion decay constant. Ambiguities in the off-shell 
amplitudes due to different parameterizations of the pion field
can be fixed at the tree level using the Adler self-consistency condition
[21]. Eqs.(3) - (7) are valid in tree approximation to first
order in the pion density. 

The results for $K^{0}$ and $\bar{K}^{0}$ can be obtained from  isospin
symmetry: $\delta M_{K^{\pm }}^{[0]}=\delta M_{K^{0}}^{[0]}=\delta M_{\bar{K}%
^{0}}^{[0]}=0$ and $V_{K^{+}}^{[0]}=-V_{K^{-}}^{[0]}=-V_{K^{0}}^{[0]}=V_{%
\bar{K}^{0}}^{[0]}$. Since at RHIC energies $\sqrt{s_{NN}}=200$ GeV the
charged pion number ratio $n_{\pi ^{+}}/n_{\pi ^{-}}$ is very close to unity 
\cite{thermo}, the pion isovector density and
the potential $V_{K}^{[0]}$ are negligible.

It is worthwhile to notice that the kaon dispersion law in nuclear matter
has a similar structure with a negative mass shift $\delta M_{K}^{2}<0$ and
a potential $V_{K^{+}}= - V_{K^{-}}>0$ \cite{Kaplan,Brown,texas}.

For pions the current algebra \cite{Weinberg} gives $Z_{\pi ^{\pm
}}^{-1}-1=(n_{s\pi ^{+}}+n_{s\pi ^{-}})/F^{2}$, $Z_{\pi
^{0}}^{-1}-1=2n_{s\pi ^{0}}/F^{2}$, $\delta M_{\pi ^{\pm }}^{2}=M_{\pi
}^{2}n_{s\pi ^{0}}/F^{2}$, $\delta M_{\pi ^{0}}^{2}=M_{\pi }^{2}(n_{s\pi
^{+}}+n_{s\pi ^{-}}-n_{s\pi ^{0}})/F^{2}$, $V_{\pi ^{\pm }}=\pm (n_{s\pi
^{+}}-n_{s\pi ^{-}})/F^{2}$, and $V_{\pi ^{0}}=0.$ The pion mass correction
at $\mu _{\pi }=0$ is in agreement with Ref. \cite{ga}.

To lowest order ChPT isospin symmetric pion matter does not change the
kaon dispersion law. The leading order effect appears at the one loop level.
The effective mass of a quasiparticle has then to be determined from
the relation 
$1/m^{{\rm eff}}=\partial^{2}\varepsilon ({\bf p})/\partial |{\bf p|}^{2}$ at $%
{\bf p}=0$ where $\varepsilon ({\bf p})$ is the single particle
energy. The vacuum ChPT corrections to the kaon self-energy are
absorbed by $M^2$ and the vacuum  renormalization constant of the kaon 
propagator and renormalize the scattering amplitudes (1). To lowest 
order in $\Sigma (p ^{2},E) = O(n_v)$ one obtains
\[ 
\varepsilon ({\bf p})=E_{{\bf p}}+\frac{\Sigma (M_{K}^{2},E_{{\bf p}})}{2E_{{\bf p}%
}} 
\] 
with $E_{{\bf p}}=+\sqrt{{\bf p}{}^{2}+M_{K}^{2}}.$ The mean-field
potential
\begin{equation}
V_{K}=\left. \frac{1}{2}\frac{\partial \Sigma (M_{K}^{2},E)}{\partial E}%
\right| _{E=M_{K}}  \label{VGL}
\end{equation}
entering the dispersion law (\ref{dl}) determines the first-order shift of
the self-energy operator from the kaon mass shell and leads
automatically to the correct dispersion law to 
order $O({\bf p}^{2}n_{v})$. Given that the
potential $V_{K}$ is known, the mass shift can be found from equation 
\begin{equation}
\delta M_{K}+V_{K}=\frac{\Sigma (M_{K}^2,M_K)}{2M_{K}}.  \label{dMG}
\end{equation}

The values $ \Sigma (M_{K}^2,M_K),$ $\delta M_{K}$, and $V_{K}$ can be
expressed in terms of the $s$- and $p$-wave scattering lengths and the $s$%
-wave effective ranges. Using Eqs. (\ref{amplitude}) and (\ref{SE}), we
obtain 
\begin{equation}
\Sigma (M_{K}^2,M_K)=-4\pi n_{v}\frac{M_{\pi }+M_{K}}{M_{\pi }}a_{0}^{+}
\label{sigm}
\end{equation}
and 
\begin{equation}
V_{K}=-\frac{2\pi n_{v}}{M_{\pi }+M_{K}}\left( a_{0}^{+}+2M_{\pi
}M_{K}(b_{0}^{+}+3a_{1}^{+})\right)~~ .  \label{VD}
\end{equation}
The corresponding mass shift can be found from Eq. (\ref{dMG}). The
self-energy operator (\ref{SE}) for kaons in isotopically symmetric pion
matter has the same form as for antikaons due to $C$-parity. The values $\delta M_{K}$
and $V_{K}$ are also isoscalars, and so $\delta M_{K^{+}}=\delta
M_{K^{0}}=\delta M_{K^{-}}=\delta M_{\bar{K}^{0}}$ and $%
V_{K+}=V_{K^{0}}=V_{K^{-}}=V_{\bar{K}^{0}}$, as distinct from the nuclear matter
case \cite{Kaplan,Brown,Waas,texas}. Moreover, in isotopically symmetric pion matter
$\Sigma(p^2,E) = \Sigma(p^2,-E)$. This is a consequence of the crossing symmetry
according to which $A^{\pm}(s,0,p^2) = \pm A^{\pm}(u,0,p^2)$ 
where $u = (p' - p_{\pi} )^2$.

Current algebra predicts \cite{Weinberg} $a_{0}^{+}=b_{0}^{+}+3a_{1}^{+}=0$.
The parameters $a_{0}^{+}$, $b_{0}^{+},$ and $a_{1}^{+}$ entering into
Eqs. (\ref{sigm}) - (\ref{VD}) receive in ChPT corrections to the order $p^{4}$. The
calculations of Ref. \cite{ber} give $a_{0}^{+}=(0.023\pm 0.012)/M_{\pi }$
and $b_{0}^{+}+3a_{1}^{+}=(0.054\pm 0.008)/M_{\pi }^{3}$. 
The representation (\ref{expand}) is valid when the amplitudes $A^{\pm }(s,0,p^2)
$ are expanded near threshold up to $O(s-(M_{\pi }+M_{K})^{2}) + O(p^2 - M_{K})^{2})$. 
The higher order terms in the 
expansion of the self-energy operator over the kaon momentum require 
the knowledge of the higher order
threshold parameters. The next order terms,
including higher partial waves $\ell \geq 2,$ do not affect the results (\ref
{sigm}) - (\ref{VD}). Eqs. (\ref{dMG}) - (\ref{VD}) provide thus a
complete calculation of the kaon mass shift and the mean-field potential in ChPT
to one loop, since the threshold parameters $a_{0}^{+}$, $b_{0}^{+},$
and $a_{1}^{+}$ are evaluated to one loop. In 
symmetric pion matter $\delta M_{\pi }\sim O(1/F^{2})$ whereas $\delta
M_{K}\sim O(1/F^{4})$. ChPT off-shell ambiguities beyond
the tree level have been investigated e.g. in \cite{ga,scherer} from
where one may conclude that physical observables are free from 
off-shell ambiguities at least up to the one loop level. 
To lowest order in the pion density, Eq.(2) and Eqs.(8) - (11) are
thus generaly valid.

According to ChPT the pseudoscalar meson masses increase with 
temperature as expected for collective modes. A similar effect exists in the
NJL model \cite{biaz}. With increasing temperature, the constituent quark
masses decrease, whereas the occupation numbers increase, pushing the $K$
mass up and the $\phi $ mass down. A decrease of the $\phi $-meson mass was
also predicted by Asakawa and Ko \cite{AsakawaKo}.

Eqs. (\ref{sigm}) - (\ref{VD}) are valid at $T\lesssim M_{\pi }$. The
chemical freeze-out temperature at RHIC $T\sim 170$ MeV \cite{thermo} 
is high, and so we should use a more phenomenological approach. 
We rewrite the $s$-wave parts of the
amplitudes $A^{I}(s,t)$ in terms of phase shifts $\delta
_{0}^{I}(p^{*}) = (a_{0}^{I}p^{*}+(b_{0}^{I}+\frac{2}{3}%
a_{0}^{I3})p^{*3})exp(-C_I\cdot \Phi_{2}^4) $, 
where $\Phi _{2} = \Phi _{2}(\sqrt{s}, M_{\pi },M_{K})=\pi
p^{*}/\sqrt{s}$ 
is the invariant $\pi K$ phase space, 
$C_{1/2} = 0.75$ and $C_{3/2} = 0.2$. The behavior of 
the $p$-wave is assumed to fixed by the $a_{1}^{I}$ scattering 
length and the resonance $K^{*}$. We thus make in
the amplitudes substitutions 
\begin{eqnarray}
&&a_{0}^{I}+b_{0}^{I}p^{*2} \rightarrow e^{i\delta _{0}^{I}(p^{*})}{\rm sin}%
\delta _{0}^{I}(p^{*})/p^{*}~~, \\
&&a_{1}^{1/2} \rightarrow a_{1}^{1/2}\frac{|(M_{\pi
}+M_{K})^{2}-M_{K^{*}}^{2}+iM_{K^{*}}\Gamma _{K^{*}}|}{%
s-M_{K^{*}}^{2}+iM_{K^{*}}\Gamma _{K^{*}}}.
\end{eqnarray}
The value $a_{1}^{3/2}$ is small and not modified. 
The amplitudes (12) and (13) satisfy unitarity. 
The experimental $\pi K$ scattering phases are 
then well reproduced, the low-temperature limit
matches smoothly with one-loop ChPT. The amplitudes $A^{I}(s,t)$
are expressed in terms of the amplitudes $A^{\pm }(s,t,p^2)$ as follows: 
$A^{1/2}(s,t)=$ $A^{+}(s,t,M_{K})^{2})+2A^{-}(s,t,M_{K})^{2})$ and
$A^{3/2}(s,t)=A^{+}(s,t,M_{K})^{2})-A^{-}(s,t,M_{K})^{2})$. 

The kaon self-energy at threshold, the mass shift and the mean-field potential 
are shown as a function of temperature in Fig. 1. 
Eq.(2) represents the self-energy in leading order in density which 
is the basic equation for the optical potential 
model \cite{landafshitz}. The region of its validity is 
restricted to small wavelengths as compared to the length of  
the mean free path. This condition, 
$\lambda \sim 1/p^{*}\lesssim 1/(n_{v}\sigma ^{tot}),$ 
is equivalent to $m_{K}\Gamma _{K}^{*}/p^{*2}\lesssim 1$. A  
simple estimate at $T=170$ MeV gives for the left hand 
side $0.08$, i.e. the criterion for the validity of Eq.(2) is 
satisfied.

At $T=170$ MeV, we obtain $\delta M_{K}=-33$ MeV and $V_{K}=21$ MeV. 
Thus the positive mass shift at low temperatures where ChPT is applicable
becomes negative with increasing temperature. The value of $ \Re \Sigma
(M_{K}^2,M_K)/(2M_{K})$ remains relatively small up to $T=200$ MeV. 
The analogy with the Walecka model for nucleons
is remarkable: The kaon mass shift at high temperatures
is large and negative, the mean-field potential is large and positive, their sum
is relatively small and negative. The kaons are therefore bound in pion matter
similar to nucleons in nuclear matter. The mean-field 
potential is, however, $C$-even, as distinct from the case of nucleons 
and kaons in nuclear matter \cite{Kaplan,Brown,texas} 
and the case of kaons in asymmetric pion matter, as discussed above 
(see Eq. (\ref{V0})).
\begin{figure}[h]
\unitlength1cm
\begin{picture}(8.,7.3)
\put(0.0,0.3){\makebox{\epsfig{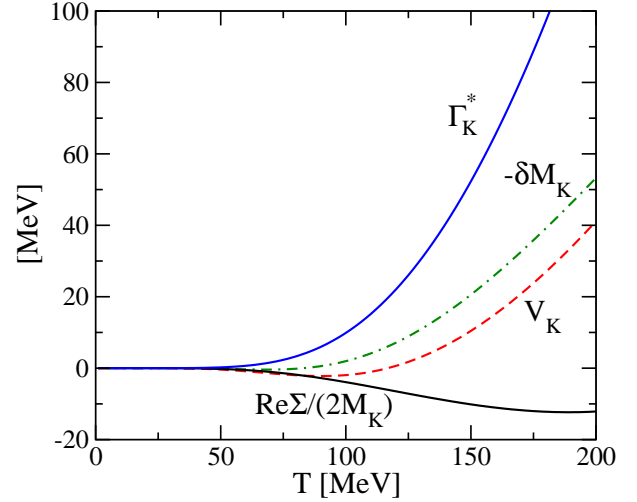}}}
\end{picture}
\caption{Self-energy $ \Re \Sigma(M^{2}_{K},M_{K} ) /(2M_K)$ of kaons at rest,
mass shift $- \delta M_K$,
mean-field potential $V_K$, and kaon collision width 
$\Gamma^{*}_K$ versus temperature
$T$ in isotopically symmetric pion matter.
}
\label{fig1}
\end{figure}
The $\phi $-meson collision broadening has been discussed in Refs. 
\cite{AsakawaKo,SeibertGale,AlvarezRusoKoch}. The physics behind this 
effect is the same as in the collision broadening of the atomic spectral
lines in hot and dense gases, discussed by Weisskopf \cite{Wei} in the
early 1930's. The $\phi \rightarrow K\bar{K}$ decay might be suppressed due to
a increasing kaon mass and/or a decreasing $\phi $-meson mass. 
Collisions of the $\phi $-meson
and the kaons keep, however, the $\phi \rightarrow K\bar{K}$ decay
channel open even at $M_{\phi }<2M_{K}$ and result in an increase of 
the total $\phi $ width, both at $M_{\phi }<2M_{K}$ and $M_{\phi }>2M_{K}$.

The kaon collision width can be found from equation
\begin{eqnarray}
\Gamma _{K}^{*}&=&\frac{1}{6M_{K}(2\pi )^{2}}\sum_{\ell }\int \left( |A_{\ell
}^{1/2}(s)|^{2}+2|A_{\ell }^{3/2}(s)|^{2}\right) 
\nonumber \\
&\times&dn_{s}\Phi _{2}(\sqrt{s}%
,M_{\pi },M_{K})  \label{gammak}
\end{eqnarray}
where $A_{\ell }^{I}(s)\ $ are the on-shell partial wave projections of the $%
\pi K$ amplitudes with total isospin $I=1/2$ and $3/2$. 
The $K^{*}$ increases the width by $\sim 40$ MeV. As a result, we get $%
\Gamma _{K}^{*}=81$ MeV at $T=170$ MeV. The kaon collision width as a
function of temperature is shown in Fig. 1.

The $\phi $-meson in-medium width can be written as 
$\Gamma _{\phi \rightarrow K\bar{K}}^{{\rm med}}$ = $\eta \Gamma _{\phi
\rightarrow K\bar{K}}^{{\rm vac}}$ where 
\begin{eqnarray}
\eta &=&\frac{1}{\pi ^{2}}\int \frac{p^{*3}(M_{\phi }^{*},m_{1}^{*},m_{2}^{*})%
}{p^{*3}(M_{\phi },M_{K},M_{K})}\frac{M_{K}^{*}\Gamma _{K}^{*}dm_{1}^{*2}}{%
(m_{1}^{*2}-M_{K}^{*2})^{2}+(M_{K}^{*}\Gamma _{K}^{*})^{2}}
\nonumber \\
&&\times \frac{M_{K}^{*}\Gamma _{K}^{*}dm_{2}^{*2}}{(m_{2}^{*2}-M_{K}^{*2})^{2}+(M_{K}^{*}%
\Gamma _{K}^{*})^{2}}.
\nonumber
\end{eqnarray}
We set $M_{\phi }^{*} = M_{\phi } - 2V_K$ and $M_{K}^{*}=M_{K}+\delta M_{K}$. 
Any modification of the real part of the kaon self-energy does not
significantly alter the in-medium $\phi$-meson decay rate as long as 
$ \Re \Sigma(M_K^2,M_K) << \Im \Sigma(M_K^2,M_K) =
M_{K}\Gamma^{*}_K$. As can be seen from Fig. 1 
this is the present case. The enhancement factor $\eta $ as a function of
temperature is shown in Fig. 2. At $T=170$ MeV, we get $\eta \sim 3$ and $%
\Gamma _{\phi }^{{\rm med}}\sim 12$ MeV. The in-medium $\phi $-meson width
increases, since the Breit-Wigner distribution allows an effective reduction
of the kaon masses and, as a consequence, an effective increase of the
available phase space.
\begin{figure}[h]
\unitlength1cm
\begin{picture}(8.,7.3)
\put(0.0,0.3){\makebox{\epsfig{file=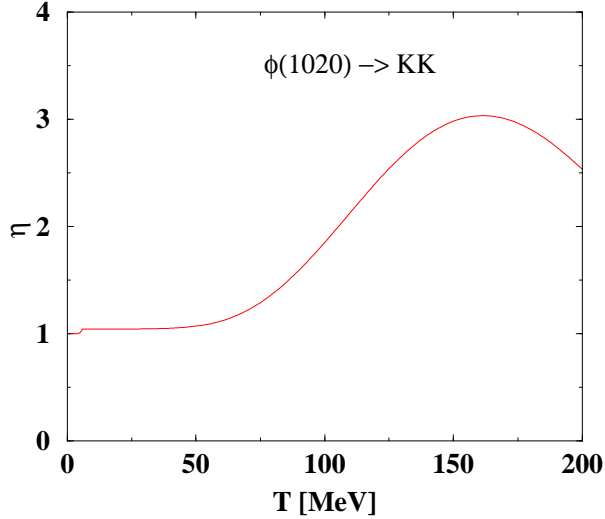,width=8.0cm}}}
\end{picture}
\caption{Enhancement factor $\eta$ of the $\phi \rightarrow K\bar{K}$ 
decay versus temperature $T$ in isotopically symmetric pion matter.
}
\label{fig2}
\end{figure}
Dileptons from $\phi $ decays leave the pion matter essentially
undistorted by final-state interactions, whereas the secondary kaons rescatter and can 
contribute to the experimental background. This can result in
an increase of the apparent dilepton branching, as one sees from the
following arguments: Let us assume that $\phi $-mesons leave the 
reaction at time $\tau $ after their creation and that kaons 
originating from in-medium $\phi $-mesons
do not rescatter. The $\phi \rightarrow K\bar{K}$
decays generate nonrelativistic $K$-mesons, in the $\phi $-meson c. m.
frame. Hence, secondary kaons move with the same velocity as $\phi $-mesons and
leave the reaction zone without rescattering with probability 
\begin{equation}
w\sim \int_{0}^{\tau }(e^{-\Gamma _{K}^{*}(\tau -t)})^{2}e^{-\Gamma _{\phi
}^{*}t}\Gamma_{\phi }~~ dt~~.  \label{w}
\end{equation}
The first term is the probability for two kaons to escape from the 
reaction zone without rescattering. The second term is the survival probability of
the $\phi$-meson at time $t$. The last term $\Gamma_{\phi}dt$ is the probability to decay
into the kaon pair during $dt$. Notice that $\Gamma_{\phi}^{*}dt$
has the meaning of a decay probability into the kaons which rescatter with pions. 
We select pairs with invariant masses of the $\phi$-meson which suffered no rescatterings.
The number of the $\phi $-mesons observed in the two-kaon channel
equals therefore 
\begin{equation}
N_{K\bar{K}}\sim e^{-\Gamma _{\phi }^{*}\tau }+
\frac{\Gamma _{\phi } }{2\Gamma
_{K}^{*}-\Gamma _{\phi }^{*}}(e^{-\Gamma _{\phi }^{*}\tau }-e^{-2\Gamma
_{K}^{*}\tau })~~. \label{NKK}
\end{equation} 
The first term arises due to the vacuum decays, whereas the second 
term is given by Eq. (\ref{w}).
The number of the 
$\phi $-mesons observed in the dilepton channel equals 
\begin{equation}
N_{e^{+}e^{-}}\sim
e^{-\Gamma _{\phi }^{*}\tau }B+(1-e^{-\Gamma _{\phi }^{*}\tau })B^{^{*}} \label{Nee}
\end{equation}
where $B\ $and $B^{*}\ $are the vacuum and in-medium dilepton branchings of
the $\phi $-mesons. We assume that the dilepton channel is
not modified, $B^{*}/B=\Gamma _{\phi }/\Gamma _{\phi }^{*}$.
The apparent dilepton branching becomes 
\begin{equation}
B^{{\rm app}}=B\frac{1+(e^{\Gamma _{\phi }^{*}\tau }-1)\Gamma _{\phi
}/\Gamma _{\phi }^{*}}{1+\frac{\Gamma _{\phi } }{2\Gamma _{K}^{*}-\Gamma
_{\phi }^{*} }(1-e^{-(2\Gamma _{K}^{*}-\Gamma _{\phi }^{*})\tau })}~~.
\label{bapp}
\end{equation}
According to transport calculations 
$e^{- \tau \Gamma _{\phi }} \sim 1/2$ at RHIC energies \cite{bravina03}. 
In such a case, varying the temperature $T = 120 \div 170$ between 
the commonly accepted values for thermal and chemical freeze-out, we get $%
B^{{\rm app}}/B\sim 2 \div 3$ which is in qualitative agreement 
with the observations from NA50, NA49, and PHENIX 
collaborations who report an  increased apparent
dilepton branching. 
Future data from RHIC and additional studies incorporating refined fireball 
dynamics as well as in-medium meson modifications might help to
completely solve the $\phi$ puzzle 
and bring more insight in the meson propagation
in a dense and hot medium.

\begin{acknowledgments}
This work is supported by
GSI (Darmstadt) under the contract T\"{U}F\"{A}ST, by RFBR grant No.
03-02-04004, DFG grant No. 436 RUS 113/721/0-1, and by Federal Program
of the Russian Ministry of Industry, Science and Technology 
No. 40.052.1.1.1112. 
\end{acknowledgments}


\end{document}